\documentclass[AMA,STIX1COL]{WileyNJD-v2}
\articletype{Research Article}%

\received{xxxx}
\revised{xxxx}
\accepted{xxxx}
\usepackage{comment}
\usepackage{rotating}
\usepackage{cases}
\usepackage{threeparttable}
\newcommand{\be}{\begin{eqnarray}}
\newcommand{\ee}{\end{eqnarray}}
\newcommand{\bee}{\begin{eqnarray*}}
\newcommand{\eee}{\end{eqnarray*}}
\newcommand{\bi}{\begin{enumerate}}
\newcommand{\ei}{\end{enumerate}}

\begin{document}

\title{Design and analysis considerations for cohort stepped wedge cluster randomized trials with a decay correlation structure}

\author[1,2]{Fan Li}
\authormark{Li}

\address[1]{\orgdiv{Department of Biostatistics}, \orgname{Yale University}, \orgaddress{\state{New Haven, Connecticut}, \country{USA}}}
\address[2]{\orgdiv{Center for Methods in Implementation and Prevention Science}, \orgname{Yale University}, \orgaddress{\state{New Haven, Connecticut}, \country{USA}}}

\corres{Fan Li, Department of Biostatistics, Yale University, New Haven, Connecticut, USA,\\
 \email{fan.f.li@yale.edu}}

\abstract[Summary]{A stepped wedge cluster randomized trial is a type of longitudinal cluster design that sequentially switches clusters to intervention over time until all clusters are treated. While the traditional posttest-only parallel design requires adjustment for a single intraclass correlation coefficient, the stepped wedge design allows multiple outcome measurements from the same cluster and so additional correlation parameters are necessary to characterize the within-cluster correlation structure. Although a number of studies have differentiated between the concepts of within-period and between-period correlations, only a few studies have allowed the between-period correlation to decay over time. In this article, we consider the proportional decay correlation structure for a cohort stepped wedge design, and provide a matrix-adjusted quasi-least squares (MAQLS) approach to accurately estimate the correlation parameters along with the marginal intervention effect. We further develop the sample size and power procedures accounting for the correlation decay, and investigate the accuracy of the power procedure with continuous outcomes in a simulation study. We show that the empirical power agrees well with the prediction even with as few as 9 clusters, when data are analyzed with MAQLS concurrently with a suitable bias-corrected sandwich variance. Two trial examples are provided to illustrate the new sample size procedure.}


\jnlcitation{\cname{%
\author{Li, F.}} (\cyear{0000}),
\ctitle{Design and analysis considerations for stepped wedge cluster randomized trials with a decay correlation structure}, \cjournal{Statistics in Medicine}, \cvol{00;00:0000--0000}.}

\maketitle

\section{Introduction}
\label{sec1}

A unique feature of cluster randomized trials (CRTs) is that intact clusters, such as schools or clinics, are randomized to intervention arms \cite{Murray1998,Donner2000}. Randomization at the cluster level often carries pragmatic considerations, for example, administrative convenience, political reasons and prevention of treatment contamination \cite{Turner2017a}. A stepped wedge cluster randomized trial is a type of longitudinal design that sequentially switches clusters to intervention during the course of the study until all clusters are treated \cite{Hussey2007}. Such designs have become increasingly popular due to their logistical flexibility and perceived ethical benefits. Because individual outcomes within the same cluster tend be more similar than those in different clusters, the intraclass correlation coefficient (ICC) plays a central role in designing CRTs. While the traditional posttest-only parallel design (i.e. parallel cluster randomized design without a baseline period as defined in Murray \cite{Murray1998}) requires adjustment for a single ICC, longitudinal cluster randomized design such as the crossover or stepped wedge design allows multiple outcome measurements from the same cluster and so naturally requires additional correlation parameters to characterize the within-cluster structure \cite{LiTurnerPreisser2018,Li2019}. Correspondingly, sample size and power calculations for stepped wedge designs necessitate the specification of more than one correlation parameters. For example, Hemming et al. \cite{Hemming2015} considered both the within-period and between-period ICCs in their sample size procedure for a cross-sectional design. Hooper et al. \cite{Hooper2016} and Li et al. \cite{LiTurnerPreisser2018} examined a three-correlation structure that additionally accounts for the within-individual repeated measurements in a closed-cohort design.

Despite existing development of multi-parameter correlation structures for designing stepped wedge trials, most of them assumed a constant between-period ICC with a few exceptions. For example, in a cross-sectional design where outcome data are obtained from a different set of participants in each cluster-period \cite{Murray1998,Barker2016}, Hemming et al. \cite{Hemming2015} allowed the between-period ICC to be different from the within-period ICC, but restricted the between-period ICC to be constant irrespective of the distance between periods. Relaxing the constant between-period ICC assumption for a cross-sectional design, Kasza et al. \cite{Kasza2017,Kasza2018} studied a non-uniform correlation structure with a decay parameter and proposed a sample size procedure that accounts for the exponential correlation decay. Grantham et al. \cite{Grantham2019} further extended their sample size procedure to allow for continuous-time correlation decay in multiple-periods CRTs with continuous recruitment. From a trial planning standpoint, if correlation decay is present, Kasza et al. \cite{Kasza2017} indicated that omitting the correlation decay in a cross-sectional design would either underestimate or overestimate the true variance of the intervention effect, which led to inaccurate sample size determination. As we demonstrate in Section \ref{sec:impact}, similar considerations carry over to closed-cohort designs, where outcome data are collected from the same set of participants in each cluster-period \cite{Murray1998,Copas2015,Barker2016}. Particularly, the constant between-period ICC assumption in Hooper et al. \cite{Hooper2016} and Li et al. \cite{LiTurnerPreisser2018} may not always be realistic and it is therefore necessary to develop alternative design and analysis strategies accounting for correlation decay in cohort stepped wedge studies.

\begin{table}[htbp]
\centering
\caption{Example correlation structures in the literature on designing stepped wedge cluster randomized trials.} \label{Table1}
\begin{tabular}{clll}
\toprule
Decay & Design feature & Correlation structure & Example references \\
\midrule
\multirow{2}{*}{No} & Cross-sectional & Nested exchangeable & Hemming et al. \cite{Hemming2015}; Hooper et al. \cite{Hooper2016}; Li et al. \cite{LiTurnerPreisser2018}\\
& Closed-cohort & Block exchangeable & Hooper et al. \cite{Hooper2016}; Li et al. \cite{LiTurnerPreisser2018,Li2018}\\
\multirow{2}{*}{Yes} & Cross-sectional & Exponential decay & Kasza et al. \cite{Kasza2017}; Kasza and Forbes \cite{Kasza2018}; Grantham et al. \cite{Grantham2019}\\
& Closed-cohort & Proportional decay & This article \\
\bottomrule
\end{tabular}
\end{table}

Two popular modeling approaches for stepped wedge designs are cluster-specific models (e.g., random-effects models) and population-averaged models \cite{Preisser2003}. The parameter estimates from a population-averaged model can be interpreted as the marginal intervention effect for the participating individuals combined over all cluster-periods, and may be preferred over the cluster-specific models for trials conducted in the health policy or health services settings \cite{LiTurnerPreisser2018}. In this article, we consider a population-averaged model with a decay correlation structure. As indicated in Table \ref{Table1}, the first contribution of this article is to fill in the gap and characterize a proportional decay structure appropriate for cohort stepped wedge designs. Such a proportional decay structure has been previously introduced in analyzing clustered longitudinal data; see, for example, Lefkopoulou et al. \cite{Lefkopoulou1989}; Shults and Morrow \cite{Shults2002} and Liu et al. \cite{Liu2002}, but has not been exemplified in CRTs with a staggered randomization. Based on the proportional decay structure, we derive a new closed-form variance expression to facilitate sample size and power determination. Based on the derived variance expression, we additionally obtain a simple-to-use design effect and study how the power depends on the correlation parameters. Since the sample size procedure requires input for the correlation parameters, accurate estimation of the correlations is instrumental for future planning of stepped wedge trials. Therefore, a second contribution of this article is to introduce a modified generalized estimating equations (GEE) approach to accurately estimate the decay correlation structure along with the marginal mean parameters. The traditional GEE \cite{Liang1986} are modified in the two ways. (i) As simple moment estimators for the decay correlation structure may not be easy to obtain, we estimate the marginal mean and correlation parameters by quasi-least squares (QLS) \cite{Chaganty1997,Chaganty1999}. The QLS approach shares the same estimating equations with GEE regarding the marginal mean parameters, but is flexible enough to accommodate non-standard correlation structures. Similar to the traditional GEE estimator, the QLS estimator is also robust to correlation misspecification. (ii) Since stepped wedge cluster randomized design frequently includes a small number of clusters, we refine the QLS approach by incorporating appropriate finite-sample bias-corrections to both the estimation of correlation parameters as well as the variance of the intervention effect.

The remainder of this article is organized as follows. The notations of cohort stepped wedge designs are introduced in Section \ref{sec:note}. In Section \ref{sec2}, we provide the proportional decay correlation structure and discuss the QLS estimators to estimate the correlation structure. In Section \ref{sec3}, we develop closed-form procedures for sample size and power calculations based on the population-averaged model coupled with the proportional decay structure. We conduct a simulation study in Section \ref{sec4} to examine the accuracy of the proposed power procedure when the trials are analyzed by the QLS approach. Section \ref{sec5} provides two illustrative examples of cohort stepped wedge designs and Section \ref{sec6} concludes.

\section{Notations and Basic Setup}\label{sec:note}
We consider a cohort stepped wedge design, where a closed cohort of individuals are enrolled at each of the $I$ participating clusters at the start of the trial. We mainly focus on cohort designs to inform the applications in Section \ref{sec5}, and will defer the discussion of cross-sectional designs to Section \ref{sec6}. We assume the trial involves a total of $T$ time periods. All clusters start from the control condition, and may be randomly chosen to switch to intervention during the course of the study, until all clusters are treated at the end of the $T$th period. Individual participants will be scheduled for outcome measurement during each period, and so each individual has a total of $T$ repeated measurements. Denote $y_{ijt}$ as the outcome for individual $j$ ($j=1,\ldots,N_i$) from cluster $i$ ($i=1,\ldots,I$) at period $t$ ($t=1,\ldots,T$). A step is defined as the pre-planned time point when at least one cluster crosses over from control to intervention. We denote the total number of steps by $S$ ($2\leq S\leq T-1$), and assume that $m_s$ clusters cross over at step $s$ such that $\sum_{s=1}^S m_s=I$. We assume a complete design in the terminology of Hemming et al. \cite{Hemming2015} such that outcome measurements are taken for all individuals during each period. Following Woertman et al. \cite{Woertman2013}, we define the baseline measurements as those taken before any cluster is randomized to intervention, and follow-up measurements as those taken after at least one cluster is randomized to intervention. We assume there are $b\geq 1$ baseline measurements planned under the control condition, and $c_s\geq 1$ follow-up measurements planned between step $s$ and step $s+1$ (or end of study). Each measurement time point is associated with a distinct time period and the total number of periods $T=b+\sum_{s=1}^S c_s$. A standard stepped wedge design is given by $b=c_s=1$ for all $s$, and $T=S+1$ ($T\geq 3$). A schematic illustration of a standard design with $I=6$ clusters and $T=4$ periods can be found in Figure \ref{fig:swd1}. We also provide a schematic illustration of an alternative design with $I=6$ clusters and $T=6$ periods in Web Figure 1.

\begin{figure}[h]
\setlength{\unitlength}{0.14in} 
\centering 
\begin{picture}(20,15.5) 
\setlength\fboxsep{0pt}
\put(1,12){\framebox(4,1.5){$0$}}
\put(7,12){\colorbox{gray!20}{\framebox(4,1.5){$1$}}}
\put(13,12){\colorbox{gray!20}{\framebox(4,1.5){$1$}}}
\put(19,12){\colorbox{gray!20}{\framebox(4,1.5){$1$}}}
\put(1,10){\framebox(4,1.5){$0$}}
\put(7,10){\colorbox{gray!20}{\framebox(4,1.5){$1$}}}
\put(13,10){\colorbox{gray!20}{\framebox(4,1.5){$1$}}}
\put(19,10){\colorbox{gray!20}{\framebox(4,1.5){$1$}}}
\put(1,8){\framebox(4,1.5){$0$}}
\put(7,8){\framebox(4,1.5){$0$}}
\put(13,8){\colorbox{gray!20}{\framebox(4,1.5){$1$}}}
\put(19,8){\colorbox{gray!20}{\framebox(4,1.5){$1$}}}
\put(1,6){\framebox(4,1.5){$0$}}
\put(7,6){\framebox(4,1.5){$0$}}
\put(13,6){\colorbox{gray!20}{\framebox(4,1.5){$1$}}}
\put(19,6){\colorbox{gray!20}{\framebox(4,1.5){$1$}}}
\put(1,4){\framebox(4,1.5){$0$}}
\put(7,4){\framebox(4,1.5){$0$}}
\put(13,4){\framebox(4,1.5){$0$}}
\put(19,4){\colorbox{gray!20}{\framebox(4,1.5){$1$}}}
\put(1,2){\framebox(4,1.5){$0$}}
\put(7,2){\framebox(4,1.5){$0$}}
\put(13,2){\framebox(4,1.5){$0$}}
\put(19,2){\colorbox{gray!20}{\framebox(4,1.5){$1$}}}
\put(-4,2.5){cluster 6}\put(-4,4.5){cluster 5}\put(-4,6.5){cluster 4}
\put(-4,8.5){cluster 3}\put(-4,10.5){cluster 2}\put(-4,12.5){cluster 1}
\put(2,16){$t=1$}\put(8,16){$t=2$}
\put(14,16){$t=3$}\put(20,16){$t=4$}
\put(7.9,14.5){$s=1$}\put(13.9,14.5){$s=2$}\put(19.9,14.5){$s=3$}
\put(2,0.5){$b=1$}\put(7.8,0.5){$c_1=1$}
\put(13.8,0.5){$c_2=1$}\put(19.8,0.5){$c_3=1$}
\end{picture}
\caption{A schematic illustration of a standard stepped wedge design with $I=6$ clusters and $T=4$ periods. Each cell with a zero entry indicates a control cluster-period and each cell with a one entry indicates an intervention cluster-period.} 
\label{fig:swd1} 
\end{figure}

\section{Analysis considerations: Models and Statistical Inference}\label{sec2}
\subsection{Population-averaged models}\label{sec:model}
The population-averaged model relates the marginal mean, $\mu_{ijt}$, to the time trend and the intervention effect by
\begin{equation}\label{eq:glm}
g(\mu_{ijt})=\beta_t+X_{it}\delta,
\end{equation}
where $g$ is the link function and $\beta_t$ is the $t$th period effect \cite{LiTurnerPreisser2018}. Further, $X_{it}$ is the intervention status, which equals $1$ or $0$ depending on whether cluster $i$ receives intervention during period $t$, and $\delta$ describes the intervention effect on the link function scale. Model \eqref{eq:glm} can be regarded as the marginal counterpart of a number of existing random-effects models, such as those proposed in Hemming et al. \cite{Hemming2015}, Hooper et al. \cite{Hooper2016} and Kasza et al. \cite{Kasza2017,Kasza2018}. Like these random-effects models, our marginal model \eqref{eq:glm} does not specify treatment by period interaction and so $\delta$ should be interpreted as the average intervention effect across periods. We write the collection of model parameters as $\theta=(\beta_1,\ldots,\beta_T,\delta)'$, the collection of intervention status for cluster $i$ (a sequence of ones preceded by zeros) as $X_i=(X_{i1},\ldots,X_{iT})'$, and define $v(\mu_{ijt})$ as a known variance function. To allow for potential correlation decay over time, we define the proportional decay correlation structure similar to Lefkopoulou et al. \cite{Lefkopoulou1989}. Specifically, we define the within-period correlation as the correlation between outcomes for two distinct individuals from the same cluster during the same period, i.e. $\text{corr}(y_{ijt},y_{ij't})=\tau$ for $j\neq j'$. The same definition is prevalently used in the posttest-only parallel designs \cite{Murray1998}. We then assume a first-order autoregressive structure for the set of within-individual repeated measurements. Therefore the within-individual correlation, which describes the association between outcomes measured at time $t$ and $t'$ of the same individual, is $\text{corr}(y_{ijt},y_{ijt'})=\rho^{|t-t'|}$, $t\neq t'$. Finally, we define the between-period correlation as the correlation between outcome measured at time $t$ for individual $j$ and outcome measured at time $t'$ for individual $j'$, and assumes a decay structure as $\text{corr}(y_{ijt},y_{ij't'})=\tau\rho^{|t-t'|}$ for $j\neq j'$, $t\neq t'$.

\begin{table}[htbp]
\centering
\caption{Examples of the proportional decay structure for cohort designs and the exponential decay structure for cross-sectional designs. The illustration is based on a stepped wedge trial with $T=3$ periods and $N_i=2$ measurements per cluster-period. Define $y_i=(y_{i11},y_{i12},y_{i13},y_{i21},y_{i22},y_{i23})'$.} \label{Table3}
\begin{tabular}{cll}
\toprule
& Proportional decay structure & Exponential decay structure \\
\midrule
$\text{corr}(y_i)$ & $\left(\begin{array}{@{}ccc|ccc@{}}
        1 & \rho & \rho^2 & \tau & \tau\rho & \tau\rho^2\\
        \rho & 1 & \rho & \tau\rho & \tau &\tau\rho\\
        \rho^2 & \rho & 1 &  \tau\rho^2 & \tau\rho & \tau\\ \hline
        \tau & \tau\rho & \tau\rho^2 & 1 & \rho & \rho^2\\
        \tau\rho & \tau & \tau\rho & \rho & 1 &\rho\\
        \tau\rho^2 & \tau\rho & \tau &  \rho^2 & \rho & 1\\
        \end{array}\right)$ &
        $\left(\begin{array}{@{}ccc|ccc@{}}
        1 & \tau\rho & \tau\rho^2 & \tau & \tau\rho & \tau\rho^2\\
        \tau\rho & 1 & \tau\rho & \tau\rho & \tau &\tau\rho\\
        \tau\rho^2 & \tau\rho & 1 &  \tau\rho^2 & \tau\rho & \tau\\ \hline
        \tau & \tau\rho & \tau\rho^2 & 1 & \tau\rho & \tau\rho^2\\
        \tau\rho & \tau & \tau\rho & \tau\rho & 1 &\tau\rho\\
        \tau\rho^2 & \tau\rho & \tau &  \tau\rho^2 & \tau\rho & 1\\
        \end{array}\right)$\\
\bottomrule
\end{tabular}
\end{table}

In a closed-cohort design, typically the between-period correlation is much smaller than the within-individual correlation for any two fixed periods, since the former is defined for measurements from two distinct individuals while the latter is defined for those from the same individual. This ordering of magnitude is reflected in the proportional decay structure because $\tau<1$ and $\text{corr}(y_{ijt},y_{ij't'})=\tau\rho^{|t-t'|}<\text{corr}(y_{ijt},y_{ijt'})=\rho^{|t-t'|}$. By comparison, the exponential decay structure by Kasza et al. \cite{Kasza2017} is based on cross-sectional designs and therefore obviates the need for modeling the within-individual correlation from repeated measurements. In fact, the exponential decay structure assumes $\text{corr}(y_{ijt},y_{ij't'})=\tau\rho^{|t-t'|}$ irrespective of whether $j=j'$ since two different sets of participants are included in two different periods for a cross-sectional design. In this respect, parameters $\tau$ and $\rho$ have similar interpretations in both structures. We provide a visual comparison of these two structures in Table \ref{Table3}. In summary, the proportional decay correlation structure is defined through two parameters,  $\tau$ and $\rho$, with the former resembling the traditional ICC definition in a parallel design and the latter controlling for the degree of correlation decay. Of note, Shults and Morrow \cite{Shults2002} and Liu et al. \cite{Liu2002} also adopted the same proportional decay structure in longitudinal CRTs with a parallel assignment, and we extend the discussion of this decay structure to CRTs with a staggered assignment.

\subsection{Quasi-least squares analysis}
We use the QLS approach introduced by Shults and Morrow \cite{Shults2002} to simultaneously estimate the intervention effect in model \eqref{eq:glm} and the correlation parameters in a cohort stepped wedge design. In particular, the QLS approach and the traditional GEE approach share the same estimating equations for the marginal mean parameters, whereas the former provides a flexible and convenient way to estimate non-standard correlation structures. Furthermore, both the QLS and GEE approaches are robust to correlation misspecification, namely estimators for the marginal intervention effect remain consistent even if the working correlation model is misspecified. In sufficiently large samples (usually $I\geq 30$), the robust sandwich variance could also be used to adequately quantify the uncertainty of the intervention effect estimate even under correlation misspecification. We refer the readers to the textbook of Shults and Hilbe \cite{Shults2014} for a full exposition on the advantages of QLS over the traditional GEE.

Write $y_{ij}=(y_{ij1},\ldots,y_{ijT})'$, $\mu_{ij}=(\mu_{ij1},\ldots,\mu_{ijT})'$, $y_i=(y_{i1}',\ldots,y_{iN_i}')'$ and $\mu_i=(\mu_{i1}',\ldots,\mu_{iN_i}')'$. Further define $D_i(\theta)=\partial\mu_i/\partial\theta'$, and let the working covariance of $y_i$ be $V_i=\phi A_i^{1/2}(\theta)R_i(\alpha_0,\alpha_1)A_i^{1/2}(\theta)$, where $\phi>0$ is the dispersion parameter, $A_i(\theta)=\text{diag}\{A_{i1}(\theta),\ldots,A_{iN_i}(\theta)\}$, $A_{ij}(\theta)=\text{diag}\{v(\mu_{ij1}),\ldots,v(\mu_{ijT})\}$, and $R_i(\alpha_0,\alpha_1)$ is a positive definite working correlation parameterized by $\alpha_0$ and $\alpha_1$. We assume the true correlation structure among elements of $y_i$ is the proportional decay structure, denoted by $R_i(\tau,\rho)$. In matrix notations, we can verify that the proportional decay structure induces separability between $\tau$ and $\rho$ in that $R_i(\tau,\rho)=G_i(\tau)\otimes F(\rho)$, where $G_i(\tau)$ is a $N_i\times N_i$ exchangeable correlation matrix (with respect to $\tau$) and $F(\rho)$ is a $T\times T$ first-order autoregressive correlation matrix, as given below
\bee
G_i(\tau)=\left(
  \begin{array}{ccccc}
    1 & \tau & \tau & \ldots & \tau \\
    \tau & 1 & \tau & \ldots & \tau \\
    \vdots & \vdots & \vdots & \ddots & \vdots \\
    \tau & \tau & \tau & \ldots & 1 \\
  \end{array}
\right),~~~~~~~~~~~F(\rho)=\left(
  \begin{array}{ccccc}
    1 & \rho & \rho^2 & \ldots & \rho^{T-1} \\
    \rho & 1 & \rho & \ldots & \rho^{T-2} \\
    \vdots & \vdots & \vdots & \ddots & \vdots \\
    \rho^{T-1} & \rho^{T-2} & \rho^{T-3} & \ldots & 1 \\
  \end{array}
\right).
\eee
We could verify that the determinant $$\det\{R_i(\tau,\rho)\}=\det\{G_i(\tau)\}^T\det\{F(\rho)\}^{N_i}=
(1-\tau)^{T(N_i-1)}\{1+(N_i-1)\tau\}^T(1-\rho^2)^{(T-1)N_i}.$$
Therefore, valid correlation values that ensure positive definite $R_i(\tau,\rho)$ should be contained in the triangular region
\be\label{eq:range}
\mathcal{S}=\left\{(\tau,\rho):-\frac{1}{\max\{N_1,\ldots,N_I\}-1}<\tau<1,-1<\rho< 1\right\}.
\ee
Finally, the inverse of the $R_i$ also exists in closed form and is given by $R_i^{-1}(\tau,\rho)=G_i^{-1}(\tau)\otimes F_i^{-1}(\rho)$, where
$$G_i^{-1}(\tau)=\frac{1}{1-\tau}I_i
-\frac{\tau}{(1-\tau)\{(1+(N_i-1)\tau)\}}J_i,~~~ F^{-1}(\rho)=\frac{1}{1-\rho^2}\{I+\rho^2 C_2-\rho C_1\},$$
$I_i$ is a $N_i\times N_i$ identity matrix, $J_i$ is a $N_i\times N_i$ matrix of ones, $C_2=\text{diag}(0,1,\ldots,1,0)$, and $C_1$ is a $T\times T$ tridiagonal matrix with zeros on the main diagonal and ones on the two sub-diagonals.

To introduce the QLS estimating equations, we further define $r_{ij}(\theta)=A_{ij}^{-1/2}(\theta)(y_{ij}-\mu_{ij})$, and write $r_i(\theta)=(r_{i1}'(\theta),\ldots,r_{iN_i}'(\theta))'$. The first-stage QLS estimates for $\theta$, $\alpha_0$ and $\alpha_1$ are obtained by alternating between the following estimating equations until convergence
\begin{eqnarray}
&&\sum_{i=1}^I D_i'(\theta)A_i^{-1/2}(\theta)R_i^{-1}(\alpha_0,\alpha_1)r_i(\theta)=0,\label{eq:gee}\\
&&\sum_{i=1}^I \frac{\partial}{\partial\alpha_0} \left\{r_i'(\theta)
R_i^{-1}(\alpha_0,\alpha_1)r_i(\theta)\right\}=0,\label{eq:alpha0}\\
&&\sum_{i=1}^I \frac{\partial}{\partial\alpha_1}\left\{r_i'(\theta)
R_i^{-1}(\alpha_0,\alpha_1)r_i(\theta)\right\}=0.\label{eq:alpha1}
\end{eqnarray}
In particular, \eqref{eq:gee} is the usual GEE coupled with the proportional decay structure, and \eqref{eq:alpha0}, \eqref{eq:alpha1} are scalar equations for the first-stage correlation estimates. Further, closed-form solutions exist for $\hat{\alpha}_0$ and $\hat{\alpha}_1$ (within an iterative step) and are provided in Web Appendix A. Chaganty and Shults \cite{Chaganty1999} showed that $\hat{\alpha}_0$ and $\hat{\alpha}_1$ are asymptotically biased for $\tau$ and $\rho$. To eliminate the large-sample bias in the first-stage correlation estimates, Chaganty and Shults \cite{Chaganty1999} provided the following second-stage estimating equations to obtain $\hat{\tau}$, $\hat{\rho}$
\begin{eqnarray}
\sum_{i=1}^I \text{tr}\left\{\frac{\partial G_i^{-1}(\hat{\alpha}_0)}{\partial\alpha_0}G_i(\tau)\right\}&=&0\label{eq:secondalpha0}\\
\text{tr}\left\{\frac{\partial F^{-1}(\hat{\alpha}_1)}{\partial\alpha_1}F(\rho)\right\}&=&0.\label{eq:secondalpha1}
\end{eqnarray}
The closed-form solution for \eqref{eq:secondalpha0} and \eqref{eq:secondalpha1} are provided by Shults and Morrow \cite{Shults2002} as
\begin{equation*}
\hat{\tau}=\sum_{i=1}^I\frac{N_i(N_i-1)\hat{\alpha}_0\{2+(N_i-2)\hat{\alpha}_0\}}
{\{1+(N_i-1)\hat{\alpha}_0\}^2}\Big/\sum_{i=1}^I\frac{N_i(N_i-1)\{1+(N_i-1)\hat{\alpha}_0^2\}}
{\{1+(N_i-1)\hat{\alpha}_0\}^2},
\end{equation*}
and $\hat{\rho}=2\hat{\alpha}_1/(1+\hat{\alpha}_1^2)$.

\subsection{Bias-corrected correlation estimation}
Although the correlation estimates obtained from the second-stage QLS estimating equations are unbiased in large samples, they could be subject to finite-sample bias. The finite-sample bias is a typical consideration in stepped wedge cluster randomized trials \cite{LiTurnerPreisser2018}, as they usually include a small number of clusters ($I\leq 30$). We refine the QLS approach with finite-sample bias-corrections to the correlation estimating equations by utilizing the cluster-leverage \cite{Preisser1996}, defined as $H_i=D_i(\theta)(\sum_{i=1}^I D_i'(\theta)V_iD_i(\theta))^{-1}D_i'(\theta)V_i$. Specifically when $I$ is small, the estimated residual $y_i-\hat{\mu}_i$ tends to be biased towards zero, and following Preisser et al. \cite{Preisser2008}, we have $E[(y_i-\hat{\mu}_i)(y_i-\hat{\mu}_i)']\approx (I-H_i)\text{cov}(y_i)=\phi(I-H_i)A_i^{1/2}\text{corr}(y_i)A_i^{1/2}$ and therefore $E[r_i(\hat{\theta})r_i'(\hat{\theta})]\approx \phi A_i^{-1/2}(I-H_i)A_i^{1/2}\text{corr}(y_i)$. This last equation suggests that $\phi^{-1}A_i^{-1/2}(I-H_i)^{-1}A_i^{1/2}r_i(\hat{\theta})r_i'(\hat{\theta})$ is a better estimator for $\text{corr}(y_i)$ compared to the simple cross-product $\phi^{-1}r_i(\hat{\theta})r_i'(\hat{\theta})$, since the former accounts for finite-sample bias in a multiplicative fashion. Further observe that
\bee
r_i'(\theta)
R_i^{-1}(\alpha_0,\alpha_1)r_i(\theta)&=&\text{tr}\{R_i^{-1}(\alpha_0,\alpha_1)r_i(\theta)r_i'(\theta)\}
\propto \phi^{-1}\text{tr}\{R_i^{-1}(\alpha_0,\alpha_1)r_i(\theta)r_i'(\theta)\}
\eee
for all values of $\alpha_0$, $\alpha_1$ and $\theta$, we propose to replace the first-stage estimating equations \eqref{eq:alpha0} and \eqref{eq:alpha1} by
\begin{eqnarray}
&&\sum_{i=1}^I \frac{\partial}{\partial\alpha_0} \text{tr}\left\{
R_i^{-1}(\alpha_0,\alpha_1)\tilde{R}_i(\theta)\right\}
=\sum_{i=1}^I \frac{\partial}{\partial\alpha_0} \text{tr}\left[
\left\{G_i^{-1}(\alpha_0)\otimes F^{-1}(\alpha_1)\right\}\tilde{R}_i(\theta)\right]=0\label{eq:refalpha0}\\
&&\sum_{i=1}^I \frac{\partial}{\partial\alpha_1} \text{tr}\left\{
R_i^{-1}(\alpha_0,\alpha_1)\tilde{R}_i(\theta)\right\}=
\sum_{i=1}^I \frac{\partial}{\partial\alpha_1} \text{tr}\left[
\left\{G_i^{-1}(\alpha_0)\otimes F^{-1}(\alpha_1)\right\}\tilde{R}_i(\theta)\right]=0\label{eq:refalpha1}
\end{eqnarray}
where $\tilde{R}_i(\theta)=\phi^{-1}A_i^{-1/2}(I-H_i)^{-1}A_i^{1/2}r_i(\theta)r_i'(\theta)$ represents the matrix-adjusted estimator for the correlation structure. The solutions obtained from \eqref{eq:refalpha0} and \eqref{eq:refalpha1} could effectively reduce the finite-sample bias in $\hat{\alpha}_0$ and $\hat{\alpha}_1$, which would in turn decrease the finite-sample bias in the QLS estimators for $\tau$ and $\rho$. Of note, similar finite-sample matrix adjustment was developed by Preisser et al. \cite{Preisser2008} for the Prentice-type GEE \cite{Prentice1988}, and we extend this finite-sample bias-correction approach to the QLS estimating equations. Accurately estimating the correlation parameters in the analysis stage has practical implications since these estimates could be used to guide the planning of future trials \cite{Murray1998}. Additional details of the matrix-adjusted estimating equations, \eqref{eq:refalpha0} and \eqref{eq:refalpha1}, along with the closed-form updates are provided in Web Appendix B.

\subsection{Bias-corrected covariance estimation}

The availability of a small number of clusters may also have implications for estimating the variance using GEE-based approaches \citep{Turner2017b}. In general, the variance of the marginal mean model parameter $\hat{\theta}$ can be estimated using the model-based variance $\Omega_1^{-1}=(\sum_{i=1}^I D_i'V_iD_i)^{-1}$ or the sandwich variance $\Omega_1^{-1}\Omega_0\Omega_1^{-1}$, where
\begin{equation}\label{eq:sand}
\Omega_0=\sum_{i=1}^I C_iD_i'V_i^{-1}B_i(y_i-\mu_i)(y_i-\mu_i)'B_i'V_i^{-1}D_iC_i,
\end{equation}
and both ${\Omega}_0$, ${\Omega}_1$ are evaluated at $(\hat{\theta},\hat{\tau},\hat{\rho})$. When both $C_i$ and $B_i$ are identity matrices, equation (\ref{eq:sand}) reduces to the uncorrected sandwich estimator of Liang and Zeger \cite{Liang1986}, which we denote as BC0. BC0 provides valid inference regardless of the correct specification of the working correlation $R_i$, as long as the number of clusters is sufficiently large ($I\geq 30$), while the consistency of the model-based variance requires the correct specification of the correlation structure. As $r_i(\hat{\theta})$ is biased towards zero with a limited number of clusters, BC0 is likely to underestimate the variance and alternative choices of matrices $C_i$ and $B_i$ may be necessary to provide a partial correction to the finite-sample bias \cite{Turner2017b}. We consider three popular approaches for bias-corrections summarized below and also in Table \ref{Table4}: the finite-sample correction due to Kauermann and Carroll \cite{Kauermann2001}, or BC1, is given by $C_i=I$ and $B_i=(I-H_i)^{-1/2}$; the finite-sample correction due to Mancl and DeRouen \cite{Mancl2001}, or BC2, is given by $C_i=I$ and $B_i=(I-H_i)^{-1}$; the finite-sample correction due to Fay and  Graubard \cite{Fay2001}, or BC3, given by $C_i=\text{diag}\{(1-\min\{\zeta,[{D_i'V^{-1}_iD_i\Omega_1^{-1}}]_{jj}\})^{-1/2}\}$ and $B_i=I$, where the bound parameter $\zeta$ is a user-defined constant ($<1$) with a default value $0.75$. Because the matrix elements of the cluster leverage are between $0$ and $1$, we generally have $\text{BC0}<\text{BC1}<\text{BC2}$ in terms of amount of correction \cite{Preisser2008}. Further, Scott et al. \cite{Scott2014} have shown that BC3 tends to be close to BC1. Of note, the estimation of dispersion parameter should only affect the model-based variance. Similar to Liang and Zeger \cite{Liang1986}, we propose to consistently update the dispersion parameter from iteration $s$ to $s+1$ by $\hat{\phi}^{(s+1)}=\hat{\phi}^{(s)}{\sum_{i=1}^I \text{tr}(\tilde{R}_i)}/\{\sum_{i=1}^I TN_i-(T+1)\}$.

\begin{table}[htbp]
\centering
\caption{Summary of bias-corrected sandwich variance estimators for $\hat{\theta}$.} \label{Table4}
\begin{tabular}{llccl}
\toprule
 Label & Correction & $C_i$ & $B_i$ & References \\ \midrule
BC0 & none & $I$ & $I$ & Liang and Zeger\cite{Liang1986} \\
BC1 & less & $I$ & $(I-H_i)^{-1/2}$ & Kauermann and Carroll\cite{Kauermann2001} \\
BC2 & more & $I$ & $(I-H_i)^{-1}$ & Mancl and DeRouen\cite{Mancl2001} \\
BC3 & less & $\text{diag}\{(1-\min\{\zeta,[{D_i'V^{-1}_iD_i\Omega_1^{-1}}]_{jj}\})^{-1/2}\}$
& $I$ & Fay and Graubard\cite{Fay2001} \\
\bottomrule
\end{tabular}
\end{table}

\section{Design Considerations: Sample Size and Power Calculations}
\label{sec3}

\subsection{Closed-form variance of the intervention effect}
Under the null hypothesis $H_0$: $\delta=\delta_0$, the large-sample variance of $\sqrt{n}(\hat{\delta}-\delta_0)$ is provided by the $(T+1,T+1)$ element of the large-sample covariance matrix of $\sqrt{n}(\hat{\theta}-\theta_0)$. Since the QLS estimator $\hat{\delta}$ is asymptotically normal, we could use the $z$-test statistic $\hat{\delta}/\sqrt{\text{var}(\hat{\delta})}$ to test the null of no intervention effect $H_0$: $\delta=0$, and the power to detect an intervention effect of size $\delta\neq 0$ with a prescribed type I error rate $\alpha$ is approximately
\begin{equation}\label{eq:powerz}
\text{power}=\Phi\left(z_{\alpha/2}+\frac{|\delta|}{\sqrt{\text{var}(\hat{\delta})}}\right),
\end{equation}
where $\Phi$ is the standard normal cumulative distribution function and $z_{\alpha/2}$ is the normal quantile such that $\Phi(z_{\alpha/2})=1-\alpha/2$. Because there is uncertainty in estimating the asymptotic variance $\text{var}(\hat{\delta})$, an alternative two-sided test uses the same statistic but refers to the $t$-distribution. We consider two choices of degrees of freedom (DoF). The first DoF dates back to Mancl and DeRouen \cite{Mancl2001} and equals the number of clusters minus the number of regression parameters; this DoF has been previously used in the GEE analyses of parallel CRTs \cite{Li2017}, three-level CRTs\cite{Teerenstra2010}, crossover CRTs\cite{Li2019}, stepped wedge CRTs\cite{LiTurnerPreisser2018,Grayling2017a}, and shown to have test size not exceeding the nominal level. The second DoF was suggested in the PhD dissertation of Ford \cite{Ford2018} and specifies $\text{DoF}=I-2$. This choice of DoF was found to provide excellent control of type I error rate for GEE analyses of both parallel CRTs and stepped wedge CRTs in Ford and Westgate \cite{Ford2017,Ford2018}. With the same effect size $\delta$ and prescribed type I error rate $\alpha$, the power of the $t$-test is approximately
\begin{equation}\label{eq:powert}
\text{power}=\Psi_{\text{DoF}}\left(t_{\alpha/2,\text{DoF}}+\frac{|\delta|}{\sqrt{\text{var}(\hat{\delta})}}\right),
\end{equation}
where $\Psi_{\text{DoF}}$ is the $t$ distribution function and the quantile $t_{\alpha/2}$ is chosen such that $\Psi_{\text{DoF}}(t_{\alpha/2})=1-\alpha/2$. We notice that because the $t$-distribution has a heavier tail compared with the standard normal distribution, the QLS $z$-test is more likely to result in a inflated type I error rate with the use of BC0 than the corresponding QLS $t$-test. As the bias-corrected sandwich variance estimators (BC1, BC2, and BC3) provide different degrees of inflation relative to the uncorrected variance BC0, an investigation of both the $z$- and $t$-tests coupled with the collection of alternative variance estimators could help inform the practical choice among the analytical options for stepped wedge CRTs.

To assist the design of stepped wedge trials allowing for correlation decay, we derive a new closed-form variance expression for $\hat{\delta}$ assuming the outcome is continuous and $g$ is the identity link function. We will return to categorical outcomes and nonlinear link functions in Section \ref{sec6}. To do so, we follow Shih \citep{Shih1997} and assume the covariance of $Y_i$ to be known as $\text{var}(Y_i)=V_i$. Therefore, $\text{var}(\hat{\delta})$ is the $(T+1,T+1)$ element of the model-based variance ${\Omega}_1^{-1}$. We further assume a balanced design such that an equal number of participants will be recruited in each cluster prior to the first period, so that $N_i=N$. Such a simplification assumption is routinely made in designing stepped wedge trials.

Under a balanced design, we could write the design matrix corresponding to cluster $i$ as $Z_i=1_N\otimes (I_T, X_i)$, where $1_N$ is a $N$-vector of ones. Then the variance of the intervention effect $\hat{\delta}$ equals to the lower-right element of $\phi\{\sum_{i=1}^I Z_i'R_i^{-1}(\tau,\rho)Z_i\}^{-1}$, where $\phi$ is the marginal variance. We show in Web Appendix C that a closed-form variance expression for $\hat{\delta}$ is
\begin{equation}\label{eq:vardelta}
\text{var}(\hat{\delta})=\frac{(\phi I/N)(1-\rho^2)\{1+(N-1)\tau\}}{(IU-W)(1+\rho^2)-2(IV-Q)\rho},
\end{equation}
where the design constant $U=\sum_{i=1}^I\sum_{t=1}^{T}X_{it}$ is the total number of cluster-periods exposed under the intervention condition, $W=\sum_{t=1}^{T}\left(\sum_{i=1}^I X_{it}\right)^2$ is the squared number of clusters receiving the intervention summed across periods, $V=\sum_{i=1}^I\sum_{t=1}^{T-1}X_{it}X_{i,t+1}$ and $Q=\sum_{t=1}^{T-1}\left(\sum_{i=1}^I X_{it}\right)\left(\sum_{i=1}^I X_{i,t+1}\right)$ are cross-product terms resulting from the decay correlation structure. It is interesting to see that this variance expression does not depend on the magnitude of the period effect $\beta_t$ as long as they are controlled for in the marginal mean model. Noticeably, the QLS-based variance \eqref{eq:vardelta} extends the formula due to Liu et al. \cite{Liu2002} to longitudinal cluster designs with staggered randomization. Further, as the cohort size $N$ becomes large, the variance expression converges to
\begin{equation}\label{eq:varlim}
\lim_{N\rightarrow\infty} \text{var}(\hat{\delta})=\frac{\phi I(1-\rho^2)\tau}{(IU-W)(1+\rho^2)-2(IV-Q)\rho},
\end{equation}
which is a finite constant since $|\rho|<1$ and $\tau>0$ for large $N$, according to \eqref{eq:range}. Therefore, the limit of the variance is a positive constant determined by available design resources $I$, $T$ and two correlation values $\tau$ ,$\rho$, and cannot be made arbitrarily small. In other words, the power of the stepped wedge design may not be increased to one by solely increasing the cohort size, which is consistent with the known results for parallel cluster randomized designs \citep{Murray1998}. For this reason, when $N$ is large, variance \eqref{eq:varlim} could be used in the design stage to approximate the variance \eqref{eq:vardelta}. Finally, given hypothesized values for $I$, $N$, $T$ and correlation parameters, variance expression \eqref{eq:vardelta} or \eqref{eq:varlim} can be used in equations \eqref{eq:powerz} and \eqref{eq:powert} to obtain the predicted power.

\subsection{The design effect}
For determining the required sample size based on equation \eqref{eq:powerz} and \eqref{eq:powert}, it is straightforward to solve $N$ by fixing the required number of clusters $I$ but not the other way around. However, with a pre-determined cohort size $N$ for each cluster, we could postulate a series of values for $I$ and find the smallest value such that the estimated power is at least equal to the prescribed level. Additionally, in the following case studied by Woertman et al. \cite{Woertman2013}, we could derive a simple expression for the design effect (DE) relative to an individually randomized trial to simplify sample size calculation. Specifically, we assume that an equal number of clusters switch to intervention at each step so that $m_s=m$, and further an equal number of measurements are taken between steps such that $c_s=s$ for all $s=1,\ldots,S$. We then write the total number of clusters $I=Sm$ and total number of periods $T=b+Sc$, and the design constants become
\begin{equation*}
\begin{split}
U&=\frac{1}{2}S(S+1)mc,~~W=\Big(\frac{1}{3}S^3+\frac{1}{2}S^2+\frac{1}{6}S\Big)m^2c,\\
V&=U-Sm,~~~~~~~~~~~~Q=W-\frac{1}{2}S(S+1)m^2.
\end{split}
\end{equation*}
Plugging the design constants back into the variance formula \eqref{eq:vardelta}, and dividing by the variance of the two-sample mean difference $4\phi/(NSm)$, we obtain
\begin{equation}\label{eq:de}
\text{DE}=\frac{3S}{2(S-1)}\frac{(1-\rho^2)}{\{(S+1)c(1-\rho)^2+6\rho\}}\{1+(N-1)\tau\}.
\end{equation}
The above design effect allows us to easily study how the design resources affect the statistical efficiency relative to individual randomization and how the correlation parameters affect the statistical power. For example, since the design effect is free of $b$, the relative design efficiency does not change according to the number of baseline periods. However, for fixed values of the correlation parameters, increasing the number of steps $S$ and number of measurements between steps $c$ decreases the design effect and increases the efficiency. On the other hand, for fixed design resources, larger values of the within-period correlation $\tau$ increases the design effect, confirming that $\tau$ functions as the traditional ICC of a parallel cluster randomized trial. By contrast, the role of correlation parameter $\rho$ is characterized by $f(p)=(1-\rho^2)/[(S+1)c(1-\rho)^2+6\rho]$, which is monotonically increasing on $(-1,r)$ and decreasing on $(r,1)$, where $$r=1+\frac{\sqrt{3}\{\sqrt{3}-\sqrt{2(S+1)c-3}\}}{(S+1)c-3}\in(0,1).$$
For convenience, we could define the decay parameter $d=1-\rho$ so that $d=0$ and $d=1$ correspond to no decay and total decay respectively. Since it is more plausible that $\rho\in(0,1)$, the above result suggests that with an increasing level of correlation decay, the design effect first increases to its largest value and then decreases, with the maximum design effect obtained at $\rho=r$. A numerical illustration of the design effect as a function of the decay parameter is provided in Web Figure 2.

\section{A Simulation Study}\label{sec4}

\subsection{Simulation design}
We carry out a simulation study (i) to compare the correlation estimators from the uncorrected QLS and the proposed matrix-adjusted QLS (MAQLS), and (ii) to evaluate the utility of the proposed power formula for QLS-based analyses of stepped wedge CRTs. For the second objective, we first determine the empirical type I error rates for the QLS-based tests coupled with alternative variance estimators, and then identify valid tests (those with a close-to-nominal type I error rate) whose empirical power corresponds well with the predicted power from the proposed formula. Findings specific to the second objective are informative for practical data analysis since we prefer tests that maintain a valid size and meanwhile demonstrate empirical power that is at least the magnitude of the analytical prediction.

\begin{table}[htbp]
\centering
\caption{Percent relative bias of the correlation parameters based on uncorrected quasi-least squares (QLS) and matrix-adjusted quasi-least squares (MAQLS) for each simulation scenario when the treatment effect is zero.}\label{Table5}
\begin{tabular}{cccrrcrrrr}
\toprule
\multicolumn{2}{c}{Correlations} & \multicolumn{1}{c}{Effect size} &
\multicolumn{3}{c}{Design resource} & \multicolumn{2}{c}{Percent bias for $\tau$} &
\multicolumn{2}{c}{Percent bias for $\rho$}\\
\midrule
$\tau$ & $\rho$ & $\delta$ & $I$ & $N$ & $T$ & QLS & MAQLS & QLS & MAQLS\\
\midrule
0.03 & 0.2 & 0 & 18 & 10 & 7 & -26.6 & 3.5 & -0.6 & 0.2 \\
  0.03 & 0.2 & 0 & 18 & 24 & 4 & -16.0 & 5.5 & -0.3 & 0.0 \\
  0.03 & 0.2 & 0 & 20 & 14 & 5 & -20.2 & 3.4 & -0.5 & 0.0 \\
  0.03 & 0.2 & 0 & 21 & 8 & 4 & -29.4 & 3.2 & -0.4 & 0.2 \\\smallskip
  0.03 & 0.2 & 0 & 15 & 8 & 4 & -41.3 & 4.7 & -0.6 & 0.3 \\
  0.03 & 0.8 & 0 & 16 & 12 & 5 & -26.2 & 4.4 & -0.6 & 0.0 \\
  0.03 & 0.8 & 0 & 24 & 7 & 5 & -27.7 & 0.6 & -0.6 & -0.1 \\
  0.03 & 0.8 & 0 & 12 & 8 & 5 & -49.4 & 8.2 & -2.1 & -0.4 \\
  0.03 & 0.8 & 0 & 12 & 5 & 4 & -74.2 & 6.6 & -2.3 & -0.5 \\\smallskip
  0.03 & 0.8 & 0 & 10 & 5 & 3 & -91.6 & -0.4 & -1.5 & -0.5 \\
  0.10 & 0.2 & 0 & 21 & 11 & 8 & -9.2 & 3.9 & -0.6 & 0.6 \\
  0.10 & 0.2 & 0 & 24 & 11 & 7 & -8.1 & 3.4 & -0.4 & 0.6 \\
  0.10 & 0.2 & 0 & 15 & 16 & 6 & -11.8 & 7.1 & -0.6 & 1.1 \\
  0.10 & 0.2 & 0 & 18 & 8 & 7 & -12.7 & 3.9 & -0.8 & 0.7 \\\smallskip
  0.10 & 0.2 & 0 & 16 & 7 & 5 & -16.7 & 3.7 & -0.7 & 1.0 \\
  0.10 & 0.8 & 0 & 20 & 18 & 5 & -8.0 & 5.0 & -0.1 & 0.2 \\
  0.10 & 0.8 & 0 & 15 & 9 & 4 & -15.2 & 3.1 & -0.2 & 0.2 \\
  0.10 & 0.8 & 0 & 10 & 20 & 3 & -16.9 & 6.5 & -0.1 & 0.3 \\
  0.10 & 0.8 & 0 & 12 & 5 & 5 & -27.0 & 2.6 & -1.5 & 0.0 \\
  0.10 & 0.8 & 0 & 9 & 7 & 4 & -29.7 & 3.8 & -1.0 & 0.1 \\
\bottomrule
\end{tabular}
\end{table}

Within-cluster correlated continuous outcomes were generated from a multivariate normal distribution with mean given by $\mu_{ijt}=\beta_t+X_{it}\delta$ and covariance $\phi R(\tau,\rho)$, where $R(\tau,\rho)$ is the proportional decay structure defined in Section \ref{sec:model}. We set the marginal variance $\phi=1$ and assumed a gently increasing period effect such that $\beta_1=0$ and $\beta_{t+1}-\beta_t=0.1\times (0.5)^{t-1}$ for $t\geq 1$. As discussed before, the predicted power should be insensitive to the magnitude of the period effects as long as they are accounted for in the QLS analyses. We fix the effect size $\delta/\phi^{1/2}$ at zero for studying test size and choose $\delta/\phi^{1/2}$ from $\{0.2,0.3,0.4,0.5\}$ for studying power. We choose the within-period correlation $\tau\in\{0.03,0.1\}$, which represent typical ICC values reported in the parallel CRTs \citep{Murray1998}. We further chose $\rho\in\{0.2,0.8\}$, representing large and moderate degree of correlation decay over time. The number of clusters are varied from $9$ to $24$ as stepped wedge CRTs usually include a limited number of clusters. We specify the number of periods $3\leq T\leq 8$ as these values are frequently reported in practice, according to recent reviews by Martin et al. \cite{Martin2016} and Grayling et al. \cite{Grayling2017b}. Finally, the cohort size are chosen as $5\leq N\leq 24$ to ensure that the predicted power is at least $80\%$. For illustration, we focus on standard stepped wedge designs so that there is only one baseline period and the number of steps $S=T-1$. In other words, an equal number of $I/S$ clusters cross over to intervention during each step, and the outcome is measured only once for each individual between consecutive steps. For each scenario, $10000$ data replications were generated and analyzed using both QLS and MAQLS. For the first objective, we report the percent relative bias in estimating $\tau$ and $\rho$. In general, an unbiased approach for estimating the correlation parameters is preferred since accurate reporting of correlations is critical for planning future trials. Web Tables 1 and 2 provide a summary of the simulation scenarios along with the convergence rates. The convergence rates are similar between QLS and MAQLS, and all exceed $97\%$. For the second objective, we consider both the $z$-tests and the $t$-tests for testing the null hypothesis of no intervention effect, coupled with five different variance estimators for $\hat{\delta}$, namely, the model-based variance, BC0, BC1, BC2 and BC3. The nominal type I error rate is held fixed at $5\%$, and we consider an empirical type I error rate between $4.5\%$ and $5.5\%$ to be acceptable based on the margin of error from a binomial model with $10000$ replications. By a similar reasoning, since the predicted power in each scenario is at least $80\%$, we consider an empirical power that differs by no more than $0.8\%$ from the predicted value to be acceptable.

\subsection{Results}

\begin{table}[htbp]
\centering
\caption{Simulation scenarios, nominal size, along with the empirical type I error rates corresponding to the MAQLS $z$-test and $t$-test ($\text{DoF}=I-2$) with different variance estimators. Empirical type I error rates between $4.5\%$ and $5.5\%$ are highlighted in boldface and considered acceptable based on the margin of error from a binomial model with $10000$ Monte Carlo replications. Pred: nominal type I error rate; MB: model-based variance; BC0: uncorrected sandwich variance of Liang and Zeger (1986); BC1: bias-corrected sandwich variance of Kauermann and Carroll (2001); BC2: bias-corrected sandwich variance of Mancl and DeRouen (2001); BC3: bias-corrected sandwich variance of Fay and Graubard (2001).}\label{tb:main1}
\begin{tabular}{cccrrrrrrrrrrrrrrr}
\toprule
& & & & & & \multicolumn{6}{c}{$z$-test} & \multicolumn{6}{c}{$t$-test}\\
 \cmidrule(lr){7-12}\cmidrule(lr){13-18}
 $\tau$ & $\rho$ & $\delta$ & $I$ & $N$ & $T$
& Pred & MB & BC0 & BC1 & BC2 & BC3 & Pred & MB & BC0 & BC1 & BC2 & BC3\\
\midrule
0.03 & 0.2 & 0 & 18 & 10 & 7 & 5.0 & \bf{4.8} & 8.4 & 6.8 & \bf{5.1} & 7.1 & 5.0 & 3.4 & 6.3 & \bf{4.7} & 3.8 & \bf{4.9} \\
0.03 & 0.2 & 0 & 18 & 24 & 4 & 5.0 & \bf{4.9} & 8.1 & 6.5 & \bf{5.0} & 6.5 & 5.0 & 3.4 & 6.0 & \bf{4.7} & 3.6 & \bf{4.7} \\
0.03 & 0.2 & 0 & 20 & 14 & 5 & 5.0 & \bf{5.1} & 8.3 & 6.9 & 5.8 & 7.0 & 5.0 & 3.8 & 6.5 & \bf{5.4} & 4.4 & \bf{5.5} \\
0.03 & 0.2 & 0 & 21 & 8 & 4 & 5.0 & \bf{5.2} & 7.9 & 6.6 & \bf{5.3} & 6.6 & 5.0 & 3.8 & 6.2 & \bf{5.0} & 4.0 & \bf{4.9} \\\smallskip
0.03 & 0.2 & 0 & 15 & 8 & 4 & 5.0 & 5.7 & 9.7 & 7.7 & 5.8 & 7.7 & 5.0 & 3.7 & 7.0 & \bf{5.3} & 4.1 & \bf{5.3} \\
0.03 & 0.8 & 0 & 16 & 12 & 5 & 5.0 & 5.6 & 9.0 & 7.3 & 5.6 & 7.3 & 5.0 & 3.8 & 6.7 & \bf{5.1} & 3.8 & \bf{5.1} \\
0.03 & 0.8 & 0 & 24 & 7 & 5 & 5.0 & \bf{5.3} & 7.7 & 6.6 & \bf{5.5} & 6.6 & 5.0 & 4.3 & 6.2 & \bf{5.2} & 4.3 & \bf{5.3} \\
0.03 & 0.8 & 0 & 12 & 8 & 5 & 5.0 & 6.0 & 10.6 & 7.9 & 5.8 & 7.9 & 5.0 & 3.5 & 7.0 & \bf{4.9} & 3.2 & \bf{5.0} \\
0.03 & 0.8 & 0 & 12 & 5 & 4 & 5.0 & 5.7 & 10.3 & 7.8 & 5.7 & 7.5 & 5.0 & 3.4 & 6.9 & \bf{5.0} & 3.5 & \bf{4.9} \\\smallskip
0.03 & 0.8 & 0 & 10 & 5 & 3 & 5.0 & 6.3 & 11.2 & 8.2 & 5.6 & 7.1 & 5.0 & 3.0 & 7.0 & \bf{4.6} & 3.1 & 3.8 \\
0.10 & 0.2 & 0 & 21 & 11 & 8 & 5.0 & \bf{5.0} & 8.1 & 6.9 & 5.7 & 7.0 & 5.0 & 3.8 & 6.6 & \bf{5.3} & 4.4 & 5.6 \\
0.10 & 0.2 & 0 & 24 & 11 & 7 & 5.0 & \bf{5.1} & 7.7 & 6.4 & \bf{5.3} & 6.5 & 5.0 & 3.8 & 6.1 & \bf{5.1} & 4.3 & \bf{5.2} \\
0.10 & 0.2 & 0 & 15 & 16 & 6 & 5.0 & \bf{4.8} & 9.7 & 7.5 & 5.7 & 7.8 & 5.0 & 3.1 & 6.8 & \bf{5.2} & 4.0 & \bf{5.3} \\
0.10 & 0.2 & 0 & 18 & 8 & 7 & 5.0 & \bf{4.8} & 8.8 & 7.1 & 5.6 & 7.3 & 5.0 & 3.4 & 6.7 & \bf{5.2} & 4.0 & \bf{5.4} \\\smallskip
0.10 & 0.2 & 0 & 16 & 7 & 5 & 5.0 & \bf{5.2} & 9.1 & 7.2 & \bf{5.5} & 7.3 & 5.0 & 3.4 & 6.6 & \bf{5.1} & 3.8 & \bf{5.2} \\
0.10 & 0.8 & 0 & 20 & 18 & 5 & 5.0 & \bf{5.3} & 7.9 & 6.4 & \bf{5.2} & 6.5 & 5.0 & 3.8 & 6.1 & \bf{4.8} & 3.9 & \bf{4.8} \\
0.10 & 0.8 & 0 & 15 & 9 & 4 & 5.0 & \bf{5.5} & 9.1 & 7.2 & \bf{5.3} & 7.0 & 5.0 & 3.7 & 6.6 & \bf{4.8} & 3.7 & \bf{4.6} \\
0.10 & 0.8 & 0 & 10 & 20 & 3 & 5.0 & 5.6 & 11.7 & 8.5 & 6.0 & 7.4 & 5.0 & 2.8 & 7.3 & \bf{4.9} & 3.0 & 4.0 \\
0.10 & 0.8 & 0 & 12 & 5 & 5 & 5.0 & 5.8 & 10.5 & 7.9 & 5.7 & 7.9 & 5.0 & 3.3 & 6.9 & \bf{4.9} & 3.3 & \bf{4.9} \\
0.10 & 0.8 & 0 & 9 & 7 & 4 & 5.0 & 6.0 & 12.0 & 8.6 & 5.7 & 8.3 & 5.0 & 2.8 & 7.2 & \bf{4.7} & 2.9 & \bf{4.5} \\
\bottomrule
\end{tabular}
\end{table}

For the first objective, we summarize in Table \ref{Table5} and Web Table 3 the percent relative bias in estimating the correlations with QLS and MAQLS. It is evident that the percent bias in estimating the within-period correlation $\tau$ is much larger than that in estimating the correlation parameter $\rho$, without respect to the incorporation of matrix adjustment to the first-stage estimating equations (\ref{eq:alpha0}) and (\ref{eq:alpha1}). However, the QLS estimator for $\tau$ exhibits noticeable downward bias, especially when the number of clusters is not large. By contrast, MAQLS substantially reduces such finite-sample bias and improves the estimation of $\tau$. On the other hand, the QLS estimator for the parameter $\rho$ seems more accurate in that the absolute percent bias only occasionally exceeds one. Nevertheless, MAQLS still mildly improves the estimation of $\rho$ in that the absolute percent bias is always maintained under one. The comparative findings between QLS and MAQLS are consistent regardless of the magnitude of intervention effect $\delta$. Therefore, MAQLS is the preferred approach because it provides much less biased estimates for the correlation parameters; these more accurate correlation estimates will eventually facilitate accurate estimation of sample size and power for future cohort stepped wedge trials.

\begin{table}[htbp]
\centering
\caption{Simulation scenarios, predicted power, along with the difference between empirical and predicted power corresponding to the MAQLS $z$-test and $t$-test ($\text{DoF}=I-2$) with different variance estimators. Differences from the prediction within $0.8\%$ are highlighted in boldface and considered acceptable. Pred: predicted power; MB: model-based variance; BC0: uncorrected sandwich variance of Liang and Zeger (1986); BC1: bias-corrected sandwich variance of Kauermann and Carroll (2001); BC2: bias-corrected sandwich variance of Mancl and DeRouen (2001); BC3: bias-corrected sandwich variance of Fay and Graubard (2001).}\label{tb:main2}
\begin{tabular}{cccrrrrrrrrrrrrrrr}
\toprule
& & & & & & \multicolumn{6}{c}{$z$-test} & \multicolumn{6}{c}{$t$-test}\\
 \cmidrule(lr){7-12}\cmidrule(lr){13-18}
 $\tau$ & $\rho$ & $\delta$ & $I$ & $N$ & $T$
& Pred & MB & BC0 & BC1 & BC2 & BC3 & Pred & MB & BC0 & BC1 & BC2 & BC3\\
\midrule
0.03 & 0.2 & 0.3 & 18 & 10 & 7 & 89.9 & \bf{-0.5} & 1.1 & -1.0 & -3.4 & -0.9 & 86.0 & \bf{0.2} & 2.4 & \bf{-0.1} & -3.4 & \bf{0.2} \\
0.03 & 0.2 & 0.3 & 18 & 24 & 4 & 88.6 & \bf{-0.4} & 1.5 & \bf{-0.4} & -2.9 & \bf{-0.5} & 84.4 & \bf{0.5} & 3.2 & \bf{0.5} & -2.5 & \bf{0.4} \\
0.03 & 0.2 & 0.3 & 20 & 14 & 5 & 89.7 & \bf{0.0} & 1.5 & \bf{-0.5} & -2.6 & \bf{-0.3} & 86.2 & \bf{0.3} & 2.6 & \bf{0.2} & -2.5 & \bf{0.4} \\
0.03 & 0.2 & 0.4 & 21 & 8 & 4 & 87.5 & \bf{-0.5} & 1.3 & \bf{-0.8} & -3.4 & -0.9 & 83.9 & \bf{-0.1} & 2.2 & \bf{-0.5} & -3.2 & \bf{-0.6} \\\smallskip
0.03 & 0.2 & 0.5 & 15 & 8 & 4 & 90.7 & -1.1 & 1.0 & -1.2 & -3.9 & -1.1 & 85.9 & \bf{-0.3} & 3.0 & \bf{-0.2} & -4.3 & \bf{-0.3} \\
0.03 & 0.8 & 0.2 & 16 & 12 & 5 & 88.6 & \bf{-0.5} & 1.2 & -1.4 & -4.4 & -1.3 & 83.8 & \bf{0.1} & 2.5 & \bf{-0.6} & -4.4 & \bf{-0.6} \\
0.03 & 0.8 & 0.2 & 24 & 7 & 5 & 88.2 & \bf{-0.5} & 1.1 & \bf{-0.6} & -2.4 & \bf{-0.7} & 85.2 & \bf{0.2} & 1.8 & \bf{0.0} & -2.3 & \bf{0.0} \\
0.03 & 0.8 & 0.3 & 12 & 8 & 5 & 94.1 & -1.1 & \bf{0.4} & -1.8 & -4.6 & -1.9 & 88.7 & \bf{0.3} & 2.7 & \bf{-0.4} & -4.8 & \bf{-0.3} \\
0.03 & 0.8 & 0.4 & 12 & 5 & 4 & 95.2 & \bf{-0.8} & \bf{0.7} & -1.2 & -3.8 & -1.5 & 90.3 & \bf{0.5} & 2.8 & \bf{-0.3} & -4.1 & \bf{-0.6} \\\smallskip
0.03 & 0.8 & 0.5 & 10 & 5 & 3 & 94.6 & \bf{-0.7} & 1.1 & -1.1 & -4.8 & -2.3 & 87.8 & \bf{0.8} & 4.3 & \bf{-0.3} & -6.0 & -2.4 \\
0.10 & 0.2 & 0.3 & 21 & 11 & 8 & 87.8 & \bf{-0.6} & 1.5 & \bf{-0.4} & -2.8 & \bf{-0.1} & 84.3 & \bf{-0.1} & 2.6 & \bf{0.1} & -2.8 & \bf{0.5} \\
0.10 & 0.2 & 0.3 & 24 & 11 & 7 & 87.8 & \bf{-0.6} & 1.2 & \bf{-0.3} & -2.5 & \bf{-0.1} & 84.8 & \bf{0.0} & 2.1 & \bf{0.2} & -2.2 & \bf{0.4} \\
0.10 & 0.2 & 0.4 & 15 & 16 & 6 & 90.6 & -1.3 & 1.2 & -1.0 & -4.3 & \bf{-0.8} & 85.8 & \bf{-0.8} & 2.9 & \bf{-0.3} & -4.1 & \bf{0.1} \\
0.10 & 0.2 & 0.4 & 18 & 8 & 7 & 91.6 & \bf{-0.1} & 1.5 & \bf{-0.4} & -2.7 & \bf{-0.1} & 87.9 & \bf{0.4} & 2.7 & \bf{0.3} & -2.5 & \bf{0.6} \\\smallskip
0.10 & 0.2 & 0.5 & 16 & 7 & 5 & 88.6 & \bf{-0.8} & 1.9 & \bf{-0.6} & -3.5 & \bf{-0.4} & 83.8 & \bf{-0.1} & 3.4 & \bf{0.3} & -3.5 & \bf{0.6} \\
0.10 & 0.8 & 0.2 & 20 & 18 & 5 & 86.1 & \bf{-0.5} & 1.8 & \bf{-0.5} & -3.0 & \bf{-0.6} & 82.1 & \bf{-0.3} & 2.7 & \bf{0.0} & -3.2 & \bf{0.0} \\
0.10 & 0.8 & 0.3 & 15 & 9 & 4 & 89.5 & -1.0 & 1.0 & -1.5 & -4.6 & -1.8 & 84.5 & \bf{-0.1} & 2.6 & \bf{-0.8} & -4.8 & -1.1 \\
0.10 & 0.8 & 0.4 & 10 & 20 & 3 & 94.4 & \bf{-0.6} & 1.6 & \bf{-0.8} & -4.3 & -2.1 & 87.5 & \bf{0.4} & 4.6 & \bf{0.6} & -5.3 & -1.8 \\
0.10 & 0.8 & 0.4 & 12 & 5 & 5 & 93.2 & \bf{-0.5} & 0.9 & -1.3 & -4.4 & -1.4 & 87.5 & \bf{0.7} & 3.2 & \bf{-0.3} & -4.8 & \bf{-0.3} \\
0.10 & 0.8 & 0.5 & 9 & 7 & 4 & 97.3 & \bf{-0.6} & \bf{0.4} & -1.1 & -4.2 & -1.4 & 91.4 & 1.1 & 3.5 & \bf{-0.2} & -5.7 & \bf{-0.7} \\
\bottomrule
\end{tabular}
\end{table}

For the second objective, we present the empirical type I error rates of the $z$-tests and $t$-tests for the QLS and MAQLS analyses in Web Tables 4-6 and Web Figure 3. Overall, we observe that the matrix adjustment to the correlation estimation mildly affects the tests with the model-based variance but has little impact on the tests with the sandwich variance. This is in accordance with Lu et al. \cite{Lu2007}, who observed the same results for the GEE analyses of pretest-posttest CRTs. Since MAQLS provides more accurate estimation of the correlations, we will focus on this approach. Table \ref{tb:main1} summarizes the empirical type I error rates of the MAQLS $z$-tests and $t$-tests with $\text{DoF}=I-2$. We leave the results for $t$-tests with $\text{DoF}=I-(T+1)$ to Web Table 6 as these tests are conservative in many cases. From Table \ref{tb:main1}, we observe that MAQLS $z$-tests are more liberal than the corresponding MAQLS $t$-tests. The type I error rates of the MAQLS $z$-tests coupled with the model-based variance or BC2 are close to nominal when $I\geq 20$, while the MAQLS $z$-tests coupled with BC0, BC1 or BC3 are always liberal. By contrast, only the MAQLS $t$-tests with BC0 remain liberal, while MAQLS $t$-tests with BC1 or BC3 maintain close-to-nominal size and MAQLS $t$-tests with model-based variance or BC2 are conservative. Overall, the $t$-tests with $\text{DoF}=I-2$ and BC1 demonstrate test sizes that consistently agree with the nominal level.

Web Tables 7-9 and Web Figure 4 present the predicted and empirical power for each simulation scenario. Because we are only interested in tests that maintain close-to-nominal sizes, we summarize in Table \ref{tb:main2} the differences between the empirical and the predicted power only for the MAQLS $z$-tests and $t$-tests with $\text{DoF}=I-2$. Among the $z$-tests, only the choice of BC2 provides substantially lower power than predicted. While the choices of model-based variance, BC1 and BC3 provide adequate power for the $z$-tests in a number of scenarios, one should be cautious in adopting these tests with a small number of clusters since they may carry an inflated test size. On the other hand, the empirical power for MAQLS $t$-tests coupled with the model-based variance, BC1 or BC3 corresponds reasonably well with the analytical prediction from the proposed formula even for as few as $9$ clusters, while the empirical power for MAQLS $t$-tests with BC2 still tends to be substantially lower than predicted. Interestingly, the MAQLS $t$-test with $\text{DoF}=I-(T+1)$ coupled with model-based variance, BC1 or BC3 also demonstrates empirical power fairly close to prediction, even though these tests are more conservative under the null. These tests may not be preferred over the $t$-tests with $\text{DoF}=I-2$ since they are less powerful.

\section{Numerical Illustrations}
\label{sec5}

\subsection{The AEP study}
We illustrate the proposed sample size procedure to design a cohort stepped wedge CRT that aims to study the effect of an exercise intervention on the physical function of patients with end-stage renal disease \cite{Bennett2013}. The intervention was an accredited exercise physiologist (AEP) coordinated resistance exercise program, offered at hemodialysis clinics to improve the quality of life for dialysis patients. During the planning phase, it was determined that $I=15$ clinics (clusters) were available, and would be randomized over $T=4$ periods evenly spaced across $48$ weeks. At baseline, no exercise programs were offered to any clinic. At week 12, 36 and 48, a random subset of $5$ clinics cross over from control to intervention. A closed cohort of patients were recruited at baseline, and would be followed up during the study period. The primary patient-level outcome was the 30-second sit-to-stand (STS) test, recording the number of times a patient could rise from and return to a seated position in a 30-second time frame. The 30-second STS test was conducted at the end of each period, resulting in $4$ outcome measurements per patient. Based on a prior study within a similar context, a conservative estimate of the effect size was given by $\delta/\phi=0.325$ \cite{Cappy1999}, and the within-period correlation was estimated to be $\tau=0.03$ \cite{Littenberg2006}. With $I=15$ clusters, the simulations suggest the MAQLS $t$-test with $\text{DoF}=I-2=13$ could maintain nominal size and adequate power; we illustrate the power calculation based on the $t$-test statistic.

Given this is a standard stepped wedge design where an equal number of clinics switch to intervention at each step, we can show that $U=IT/2$, $W=I^2T(2T-1)/\{6(T-1)\}$, $V=I(T-2)/2$ and $Q=I^2T(T-2)/\{3(T-1)\}$. The variance expression \eqref{eq:vardelta} is then simplified to
\be\label{eq:varsimp}
\text{var}(\hat{\delta})=\frac{6(\phi/N)(T-1)(1-\rho^2)\{1+(N-1)\tau\}}
{I(T-2)\{T(1-\rho)^2+6\rho\}}.
\ee
If we anticipate large correlation decay so that $\rho=0.2$, the power is estimated using equation \eqref{eq:powerz} and \eqref{eq:varsimp} to be $79.4\%$ if $N=21$ and $80.5\%$ if $N=22$. Therefore at least $N=22$ patients should be recruited in each clinic to achieve $80\%$ power under the proportional decay structure. On the other hand, we could arrive at the same results by using the design effect \eqref{eq:de}. For example, in an individual randomized study, $348$ patients would be required for the hypothesized effect size. Assuming $21$ patients will be included in each clinic, the design effect is approximately $0.92$, indicating a total of $320$ patients in approximately $15.2$ clinics would be required. Since the study affords to randomize only $15$ clinics, we increase the cohort size to $N=22$, resulting in a design effect $0.94$. Therefore, $326$ patients are required for a total of $326/22\approx 14.8$ clinics, and we conclude that $22$ patients in $15$ clinics ensured $80\%$ power.

\begin{figure}[htbp]
\centering\includegraphics[scale=0.4]{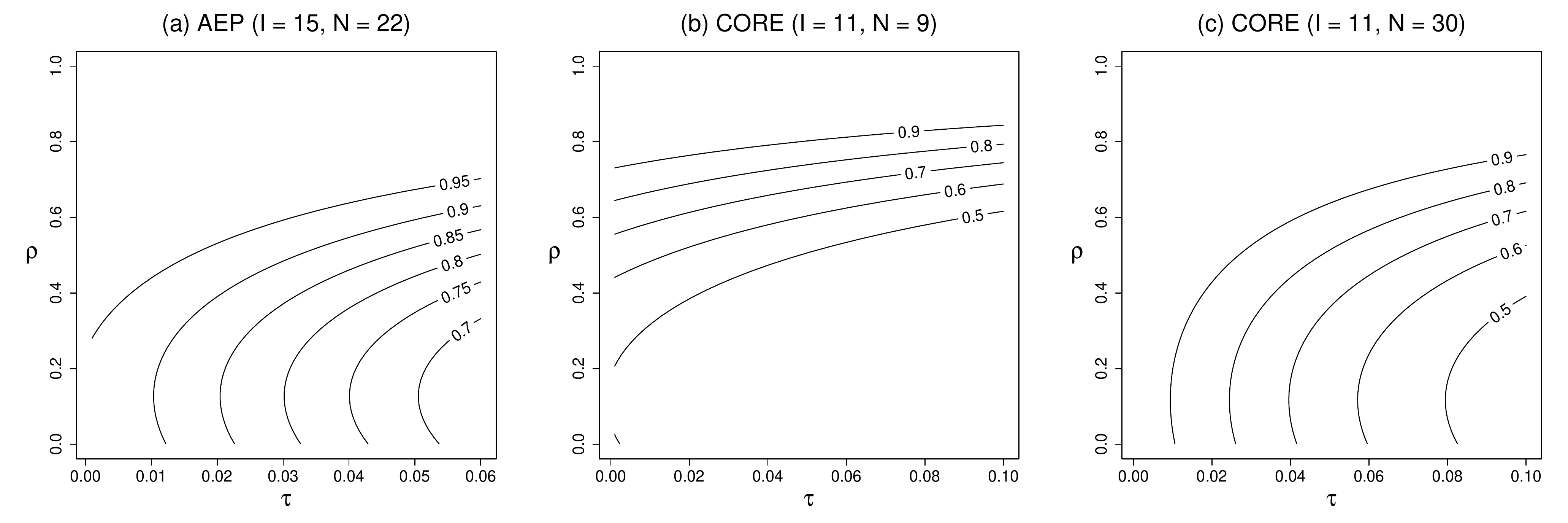}
\caption{Sensitivity analysis of study power for (a) the AEP study with $I=15$ clusters and $N=22$ individuals within each cluster, (b) the CORE study with $I=11$ clusters and $N=9$ individuals within each cluster, and (c) the CORE study with $I=11$ clusters and $N=30$ individuals within each cluster.}
\label{Fig3}
\end{figure}

While the within-period correlation estimate was available from prior studies, published estimates of the correlation parameter $\rho$ (or decay parameter $d=1-\rho$) are currently rare. For this reason, we carry out a sensitivity analysis on the power and present the results in panel (a) of Figure \ref{Fig3}, where we fix the design resources but vary $\tau\in (0.03,0.06)$ and $\rho\in (0,1)$. Note that the upper bound of the within-period correlation $\tau$ was reported by Littenberg and  MacLean \cite{Littenberg2006} and is used in this assessment. As expected, larger values of the within-period correlation reduce the study power, and further, given a certain value of the within-period correlation, a greater magnitude of decay (smaller $\rho$ or larger $d$) generally reduces the study power unless $\rho\approx 0$ (or $d\approx 1$). For the hypothesized $\tau=0.03$, the study power remains at least close to $80\%$ regardless of the correlation decay. On the other hand, the amount of correlation decay could result in further power loss if the within-period correlation $\tau$ increases. Nevertheless, the power loss is at most around $10\%$ even if the within-period correlation $\tau$ approximates the upper bound $0.06$.

\subsection{The CORE study}
We next illustrate the proposed sample size procedure by designing the CORE stepped wedge trial. The CORE study is a cluster randomized trial which aims to evaluate the patient-centered service design in health providers to improve the psychosocial recovery outcomes for people with severe mental illness in Australia \cite{Palmer2015}. The new service design intervention adopted the Experience Based Co-Design (EBCD) to identify users' positive and negative experiences of the service, and involved patients' participation to co-design solutions to the negative experiences. A total of $I=11$ teams from four health service providers would be participating the study; each team involved a number of service users who will be affected by the intervention. A stepped wedge design was considered appropriate for the study due to logistical constraint in simultaneously introducing the intervention to more than a few teams. The EBCD intervention will be delivered to the clusters in three waves, each with a duration of 9 months. Four teams will start the intervention in wave 1 and wave 2, respectively, while the remaining three teams receive the intervention in the final wave. In other words, the study includes four periods, with a baseline period lasting about 6 months.

The outcome of interest is the improvement in psychosocial recovery measured by the Recovery Assessment Scale Revised \citep[(RAS-R)][]{Lusczakoski2014}, and was measured for each user at the end of baseline period and each of the three follow-up period. The standardized effect size on the psychosocial recovery outcome was estimated to be $0.35$, and the within-period correlation was assumed to be $\tau=0.1$ \citep{Palmer2015}. Since the study affords to randomize only $11$ clusters, there may be a risk of inflated type I error rate with a $z$-test. As the $t$-test with $\text{DoF}=I-2=9$ performs best with respect to empirical size and power in the simulations, we determine the required cohort size based on a that test using expressions \eqref{eq:powert} and \eqref{eq:vardelta}. Assuming the correlation decay is only moderate so that $\rho=0.8$, power is estimated to be $0.79$ for $N=8$ and $0.81$ for $N=9$, barring drop out. Therefore $N=9$ is required to ensure $80\%$ power given a $5\%$ test size. We further conducted a sensitivity analysis to see how power changes according to the degree of correlation decay, and presented the power contour in panel (b) of Figure \ref{Fig3}. Due to the small sample size and the heavy tail of the $t$ distribution, the study is sensitive to correlation decay when $\tau=0.1$, and remains so even if $\tau$ approaches zero. On the other hand, the actual study planned to recruit $N=30$ users in each team. With this larger cohort size, panel (c) of Figure \ref{Fig3} suggests that the power becomes less sensitive to the correlation decay, especially as the within-period correlation approaches zero. For example, if $\tau\leq 0.02$, the study power remains at least around $80\%$ regardless of the amount of correlation decay.

\subsection{Consequences of specifying a non-decay correlation structure}
\label{sec:impact}
\begin{figure}[htbp]
\centering\includegraphics[scale=0.42]{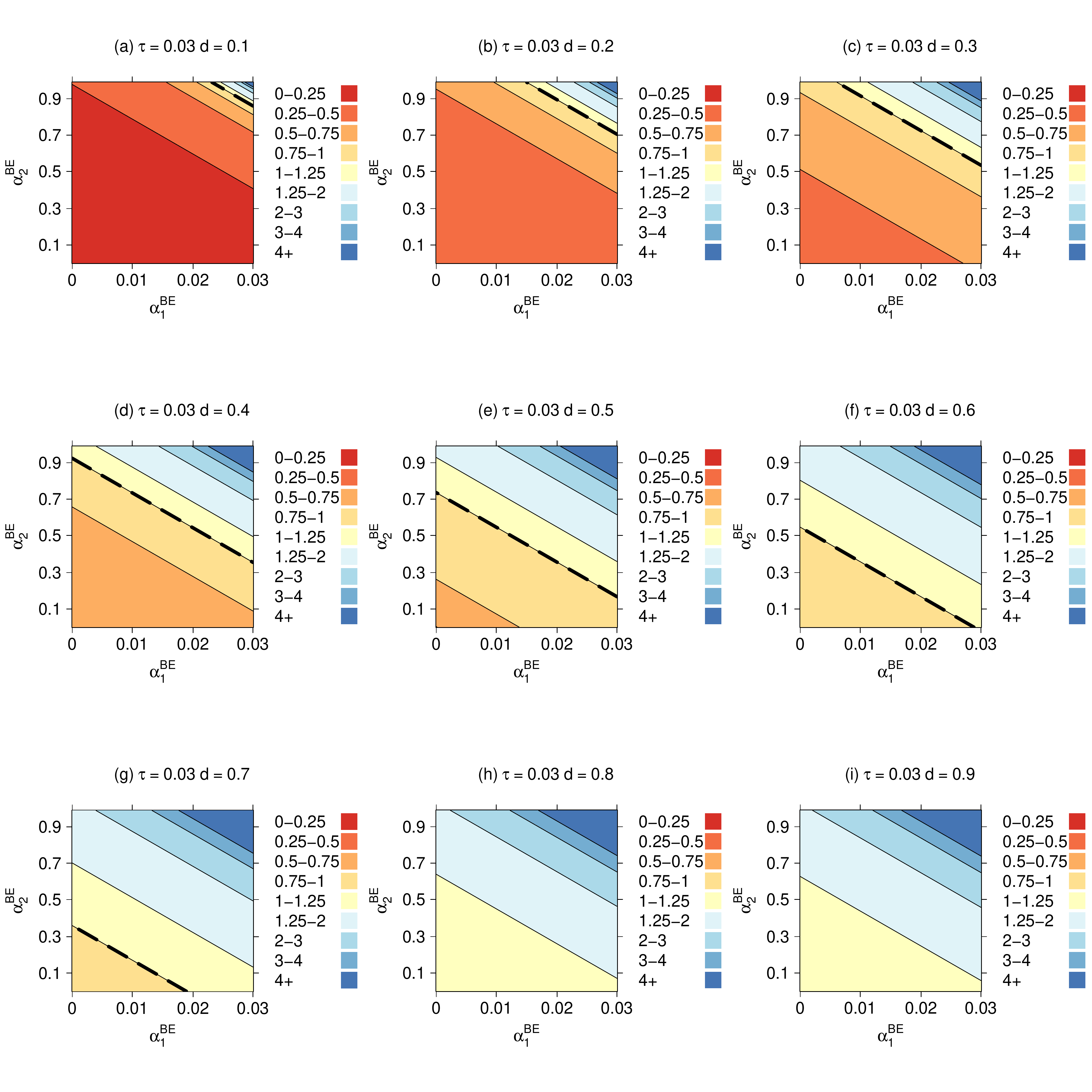}
\caption{Contour plots for the relative variance obtained under the proportional decay correlation model and the block exchangeable correlation model, for varying values of the proportional decay model decay parameter $d$ and the block exchangeable model correlation parameters $\alpha_1^{\text{BE}}$, $\alpha_2^{\text{BE}}$. In all panels, the within-period correlation $\tau=0.03$, the number of periods $T=4$ and the cohort size $N=20$. The dashed thick line indicates the equality of variances.}
\label{FigS1}
\end{figure}

For GEE analyses of cohort stepped wedge designs, Li et al. \cite{LiTurnerPreisser2018} recently developed the block exchangeable correlation structure for estimating the sample size and power. With a continuous outcome and identity link function, the block exchangeable correlation structure is also implied by the linear mixed effects model for cohort studies discussed in Hooper et al. \cite{Hooper2016}, and Girling and Hemming \cite{Girling2016}. To understand the implications of alternative correlation models, we compare the variance of the intervention effect estimator obtained under the proportional decay correlation structure to that obtained under the block exchangeable correlation structure. Specifically, both the proportional decay structure and the block exchangeable correlation structure assume a constant within-period correlation, i.e. $\text{corr}(y_{ijt},y_{ij't})=\tau$ for $j\neq j'$. However, the latter correlation model also assumes constant between-period and within-individual correlations such that $\text{corr}(y_{ijt},y_{ij't'})=\alpha_1^{\text{BE}}$ for $j\neq j'$, $t\neq t'$ and $\text{corr}(y_{ijt},y_{ijt'})=\alpha_2^{\text{BE}}$, $t\neq t'$. To focus ideas, we carry out the comparisons based on the standard stepped wedge designs with a single baseline period and an equal number of clinics switching to intervention at each step. We refer to the variance of $\hat{\delta}$ obtained under the proportional decay structure as $\text{var}^{\text{PD}}(\hat{\delta})$, whose expression is provided in \eqref{eq:varsimp}. We refer to the variance of $\hat{\delta}$ obtained under the block exchangeable structure as $\text{var}^{\text{BE}}(\hat{\delta})$, whose expression is derived in Li et al. \cite{LiTurnerPreisser2018} as
\be\label{eq:varBE}
\text{var}^{\text{BE}}(\hat{\delta})=\frac{12(\phi/N)(T-1)\lambda_3\lambda_4}
{I(T-2)\{(T-1)\lambda_3+(T+1)\lambda_4\}},
\ee
with $\lambda_3=1+(N-1)(\tau-\alpha_1^{\text{BE}})-\alpha_2^{\text{BE}}$ and $\lambda_4=1+(N-1)\tau+(T-1)(N-1)\alpha_1^{\text{BE}}+(T-1)\alpha_2^{\text{BE}}$ as the two distinct eigenvalues of the block exchangeable matrix. If we define function $h(\alpha_1^{\text{BE}},\alpha_2^{\text{BE}})=(N-1)\alpha_1^{\text{BE}}+
\alpha_2^{\text{BE}}$, we can write the relative variance as
\be\label{eq:vr}
\frac{\text{var}^{\text{PD}}(\hat{\delta})}{\text{var}^{\text{BE}}(\hat{\delta})}
=\frac{(1-\rho^2)}
{2\{T(1-\rho)^2+6\rho\}}\left[\frac{(T-1)\{1+(N-1)\tau\}}{1+(N-1)\tau+(T-1)
h(\alpha_1^{\text{BE}},\alpha_2^{\text{BE}})}+\frac{(T+1)\{1+(N-1)\tau\}}
{1+(N-1)\tau-
h(\alpha_1^{\text{BE}},\alpha_2^{\text{BE}})}\right].
\ee
For each value of the within-period correlation $\tau$ and each value of the decay parameter $d=1-\rho$ in the proportional decay model, there may exist pairs of values for $(\alpha_1^{\text{BE}},\alpha_2^{\text{BE}})$ that result in the same variance of the intervention effect. Obtaining the equal variance correspondence is equivalent to finding the straight line $h(\alpha_1^{\text{BE}},\alpha_2^{\text{BE}})=\eta$ that solves $\text{var}^{\text{PD}}(\hat{\delta})/\text{var}^{\text{BE}}(\hat{\delta})=1$. Because the relative variance is quadratic in $h(\alpha_1^{\text{BE}},\alpha_2^{\text{BE}})$, such a line may not always exist within the plausible range of $(\alpha_1^{\text{BE}},\alpha_2^{\text{BE}})$ that ensures a positive definite block exchangeable correlation matrix. We confirm this observation by plotting the relative variance for varying correlation parameters. Fixing $\tau=0.03$, $T=4$ and $N=20$ similar to the AEP study, we present in Figure \ref{FigS1} the contour of $\text{var}^{\text{PD}}(\hat{\delta})/\text{var}^{\text{BE}}(\hat{\delta})$ over the the $(\alpha_1^{\text{BE}},\alpha_2^{\text{BE}})$ plane for each level of decay $d\in\{0,1,\ldots,0.9\}$. The dashed thick line indicates the equal variance correspondence. We also present the contour plots by specifying $N=100$, $T=8$ and $\tau=0.1$ in Web Figures 5-7. It is evident that the existence and location of the equal variance line on the $(\alpha_1^{\text{BE}},\alpha_2^{\text{BE}})$ plane depends on the value of correlation decay, cohort size and number of periods. If the equal variance line exists, the value of $\tau$ and number of periods only affects its location while cohort size $N$ further affects its slope, as expected from inspecting expression \eqref{eq:vr}. Apart from the equal variance correspondence, the two variances are generally different. Depending on the gray shades (colored shades in the online version), either the proportional decay or the block exchangeable structure will lead to a larger variance of the intervention effect and require a larger sample size. As a result, using the block exchangeable correlation structure in the presence of correlation decay could lead to an overestimation or underestimation of the true intervention effect variance, and vice versa. Particularly, the variances returned from these two correlation models become close when $\alpha_1^{\text{BE}}$, $\alpha_2^{\text{BE}}$ approximate zero and when the decay parameter, $d=1-\rho$, approximates one. These two restrictions result in many nearly-zero entries in the block exchangeable and proportional decay correlation matrices, increasing the dependence of sample size estimation on the within-period correlation $\tau$. Therefore, it is anticipated that the variances from the two models become similar in this particular scenario, even though in general there could be large differences between $\text{var}^{\text{PD}}(\hat{\delta})$ and $\text{var}^{\text{BE}}(\hat{\delta})$. To summarize, our key message for cohort stepped wedge designs echo the principal findings in Kasza et al. \cite{Kasza2017} for cross-sectional designs: it is possible to grossly overestimate or underestimate the variance of the intervention effect if the correlation model is misspecified, except in restrictive scenarios. In practice, researchers could investigate the sensitivity of sample size estimates to misspecification of the correlation structure when there is limited preliminary data at the design stage of the trial.

\section{Discussion}
\label{sec6}
This article expanded on the design and analysis considerations for cohort stepped wedge CRTs in the presence of correlation decay. Since a cohort design involves repeated outcome assessments for fixed sets of individuals, we adopted the proportional decay structure of Lefkopoulou et al. \cite{Lefkopoulou1989} to characterize the within-cluster correlations among the outcome measurements. Based on a marginal mean model accounting for the treatment and period effects, we developed a new sample size and power procedure to design stepped wedge CRTs accounting for such correlation decay. To apply this procedure, a key step is to obtain reasonable values for the correlation parameters. The within-period correlation, $\tau$, is similar to the traditional ICC in a parallel cluster randomized trial, and may often be found in previous studies with a similar endpoint. By contrast, the correlation parameter, $\rho$ (or decay parameter $d=1-\rho$), is not as commonly reported in the literature, and therefore the sensitivity of power should be investigated across a range of values for $\rho$, as illustrated in Section \ref{sec5}. Given that accurate reporting of correlations is vitally important for designing future stepped wedge trials, we also provided an improved, matrix-adjusted quasi-least squares approach to estimate the correlation parameters along with the marginal mean parameters. The MAQLS has little impact on the estimation of the marginal mean parameters and the associated statistical tests coupled with the sandwich variance, but it substantially reduces the bias in estimating the within-period correlation $\tau$ and mildly improves the estimation of $\rho$, as confirmed in our simulation study.

In our simulation study with a small number of clusters, we found that, regardless of choice of variance estimators, the $t$-tests provide better control of the type I error rates compared to the $z$-tests, which are often liberal. Regardless of the two choices of DoF, the $t$-tests coupled with the model-based variance, BC1 and BC3 preserve the nominal size and demonstrate empirical power that agrees well with analytical prediction. Since the $t$ distribution with a smaller DoF has a heavier tail, which implies a smaller power under the alternative, we prefer the $t$-tests with $\text{DoF}=I-2$ for the design and analysis of cohort stepped wedge CRTs under the proportional correlation decay structure. An additional piece of evidence that favors $\text{DoF}=I-2$ is found in the recent simulation study by Ford \cite{Ford2018}, who showed that the GEE $t$-tests with $\text{DoF}=I-2$ provide satisfactory control of type I error rates even for $I=6$ but in the absence of correlation decay. Additional work is needed to investigate whether these extremely small sample sizes could provide adequate power under the proportional decay structure. On the other hand, although the $t$-tests with $\text{DoF}=I-(T+1)$ could provide adequate power using the model-based variance, BC1 or BC3, they are frequently conservative under the null with a small number of clusters. In fact, one needs a minimum of $T+2$ clusters to provide at least one DoF, which further limits the applications of such a $t$-test in the design and analysis of small stepped wedge trials.

Recent reviews of stepped wedge CRTs \cite{Barker2016,Martin2016,Grayling2017b} indicated that both the cross-sectional and cohort designs were common in practice. Although we have developed the design and analysis strategies specifically for cohort stepped wedge CRTs, a parallel discussion for cross-sectional stepped wedge CRTs could be equally informative. As discussed in Section \ref{sec:model}, the exponential decay structure is often used to model correlation decay in multi-period cluster randomized trials with repeated cross-sectional samples \cite{Kasza2017}. Assuming there are $N_i$ observations in each period for each cluster, we could write the exponential decay structure as $L_i(\tau,\rho)=(1-\tau)I_{TN_i}+\tau J_N\otimes F(\rho)$ without changing the interpretation of $\tau$ and $\rho$ from the cohort setting. To estimate the intervention effect and correlation parameters, the MAQLS approach could still be applied once we replace the second-stage estimating equations \eqref{eq:secondalpha0} and \eqref{eq:secondalpha1} by
\begin{eqnarray*}
\sum_{i=1}^I \text{tr}\left\{\frac{\partial L_i^{-1}(\hat{\alpha}_0,\hat{\alpha}_1)}{\partial\alpha_0}L_i(\tau,\rho)\right\}&=&0,\\
\sum_{i=1}^I \text{tr}\left\{\frac{\partial L_i^{-1}(\hat{\alpha}_0,\hat{\alpha}_1)}{\partial\alpha_1}L_i(\tau,\rho)\right\}&=&0.
\end{eqnarray*}
Such modifications are necessary because $\tau$ and $\rho$ are no longer separable in $L_i(\tau,\rho)$, and hence updates for $\tau$ and $\rho$ do not come in closed forms. Correspondingly, the inseparability between $\tau$ and $\rho$ also precludes the derivation of an analytical inverse $L_i^{-1}(\tau,\rho)$, and therefore one may not be able to obtain a simple algebraic expression for $\text{var}(\hat{\delta})$. As a result, sample size and power calculation requires numerically inverting the correlation matrix $L_i(\tau,\rho)$. In fact, with a continuous outcome and the identity link, it is straightforward to show that the QLS-based sample size procedure with $L_i(\tau,\rho)$ coincides with the mixed-effects model-based sample size procedure developed in Kasza et al.\cite{Kasza2017} with exponential correlation decay.

We have assumed that each individual has complete follow-up during the study. In reality, individual drop-out may be anticipated and could be accounted for in the design phase. Given an expected attrition rate $\gamma$, a simple and commonly-used strategy is to inflate the required total sample size by $1/(1-\gamma)$, so that a complete-trajectory GEE analysis may provide adequate power if the drop-out or missingness is completely at random \cite{little2002}. More sophisticated approaches that deal with monotone missingness have been studied for repeated-measure randomized trials and may be adapted to the stepped wedge context by considering staggered treatment assignments and appropriate levels of clustering \cite{Rochon1998,Jung2003}. In any event, trial implementation methodologies to prevent attrition bias in longitudinal cluster randomized trials merit further investigation \cite{Prost2015}.

One simplification we made in the sample size and power calculations was to assume equal cluster (cohort) sizes. It has been shown that cluster size imbalance leads to reduced power in parallel CRTs and therefore may be accounted for in the design phase \citep{Eldridge2006}. For a stepped wedge trial, Girling \cite{Girling2018} computed the relative efficiency of unequal versus equal cluster sizes by assuming a linear mixed-effects model without correlation decay. It was concluded that the efficiency loss due to unequal cluster sizes is unlikely to exceed $12\%$ across a wide range design of resources and correlation values. Nevertheless, a corresponding expression for the relative efficiency accounting for correlation decay is currently not available and should merit additional study. The availability of such expressions for relative efficiency could inform the amount of additional design resources required to compensate the efficiency loss due to unequal cluster sizes. Another limitation of our design strategy is that we have assumed the proportional decay correlation is the correctly specified within-cluster dependency structure. However, both the QLS or MAQLS estimators for the intervention effect remain consistent even if the correlation structure is misspecified. If it is anticipated in the design phase that the working correlation may be misspecified, one could follow the general idea of Rochon \cite{Rochon1998} to develop a modified sample size procedure based on the sandwich variance.

Finally, we have assumed a continuous outcome and an identity link function, corresponding to the scenarios of the illustrative examples. In practice, categorical outcomes could be collected in cohort stepped wedge designs. Under correlation decay, we could extend the QLS-based sample size procedure to accommodate binary and count outcomes by following the steps outlined in Section 3.2 of Li et al. \cite{LiTurnerPreisser2018}. In those cases, a further complication is that the variance is an explicit function of the marginal mean, and so the magnitude of the period effects necessarily affects the variance for the intervention effect. We plan to carry out future work to investigate the operating characteristics of such sample size procedures for binary and count outcomes.


\section*{Acknowledgments}
This work is supported within the National Institutes of Health (NIH) Health Care Systems Research Collaboratory by the NIH Common Fund through cooperative agreement U24AT009676 from the Office of Strategic Coordination within the Office of the NIH Director and cooperative agreement (UH3DA047003) from the National Institute on Drug Abuse. The content is solely the responsibility of the author and does not necessarily represent the official views of the National Institutes of Health. The author thanks Dr. John S. Preisser at the University of North Carolina at Chapel Hill for providing thoughtful comments to an earlier version of this article. The author is also grateful for the Associate Editor and two anonymous reviewers for providing helpful comments, which improved the exposition of this work.




\section*{Conflict of interest}
The author declares no potential conflict of interests.

\section*{Supporting information}
Additional supporting information including Web Appendices A--F, Web Tables and Figures, and R code for fitting the matrix-adjusted quasi-least squares with continuous outcomes may be found online in the supporting information tab for this article.

\bibliographystyle{wileyNJD-AMA}
\bibliography{SWDdecay}

\end{document}